\documentclass[12pt]{article}
\input xy
\xyoption{all}
  \usepackage{amsthm,amsfonts}
  \usepackage{amsmath}
\newcommand{\bea}{\begin{eqnarray}}
\newcommand{\eea}{\end{eqnarray}}
\begin{document}
\renewcommand{\thefootnote}{\fnsymbol{footnote}}

\thispagestyle{empty}

\title{Noncommutative oscillators from a Hopf algebra twist deformation. A first principles derivation.}
\author{P. G. Castro\thanks{{\em e-mail: pgcastro@cbpf.br}},~
B. Chakraborty\thanks{{\em e-mail: biswajit@bose.res.in}},~
R. Kullock\thanks{{\em e-mail: ricardokl@cbpf.br}}
 ~and F. Toppan\thanks{{\em e-mail: toppan@cbpf.br}}
\\ 
\\
{\it $~^{\ast}$ DM/ICE/UFJF, Campus Universit\'ario,} \\{\it cep 36036-330, Juiz de Fora (MG), Brazil.}
\\
{\it $~^\dagger$ S. N. Bose National Center for Basic Sciences,}\\ {\it JD Block, Sector III, Salt-Lake, Kolkata-700098, India.}
\\
{\it $~^{\ddagger\S}$ TEO/CBPF, Rua Dr.} {\it Xavier Sigaud 150,} \\{\it cep 22290-180, Rio de Janeiro (RJ), Brazil.}}
\maketitle
\begin{abstract}
Noncommutative oscillators are first-quantized through an abelian Drinfel'd twist deformation of a Hopf algebra and its action on a module. Several important and subtle issues making possible the quantization are solved.  The spectrum of the single-particle Hamiltonians is computed.
The multi-particle Hamiltonians are fixed, unambiguously, by the Hopf algebra coproduct.
The symmetry under particle exchange is guaranteed. In $d=2$ dimensions the rotational invariance is
preserved, while in $d=3$ the $so(3)$ rotational invariance is broken down to an $so(2)$ invariance.
\end{abstract}
\vfill
\rightline{}
\rightline{CBPF-NF-003/10}
\newpage
\section{Introduction}

In this paper the $d$-dimensional noncommutative oscillators are first-quantized in terms of an abelian Drinfel'd twist deformation of a 
Hopf algebra and its action on a module. In the literature
(for a review see \cite{{dn},{szabo},{bcgms}} and the references therein) several works have been devoted to the quantization of noncommutative dynamical systems and, in a few cases, 
of noncommutative oscillators \cite{lsz}--\cite{hok}. 
Most of the works adopt a pragmatic
viewpoint and the techniques of the noncommutative space making use of the Moyal product, see \cite{{gro},{moyal}}. On the other hand the noncommutativity can be neatly encoded \cite{{drin85},{drin88},{res}} in the framework of the Drinfel'd twist deformation of a Hopf algebra
(for a review, see \cite{{majid},{aschieri},{aschieri2},{aschieri3}}). For a noncommutativity
induced by a constant matrix the twist is abelian and the Hopf algebra under consideration is
in general given by a (deformed) Universal Enveloping Lie Algebra \cite{cct}.  There is a hyatus, however,
between the formal aspects of the Drinfel'd twist approach to noncommutativity
and its concrete application to the operatorial quantization of a dynamical system.  There are 
a few fundamental issues, not addressed in the more formal literature on Drinfel'd twist, that have to be settled and solved before applying the twist-deformed Hopf algebra framework and its action on a module to the quantization of a given dynamical system. We can
just mention some of the aspects that have been discussed (and solved) in previous works of our collaboration (with different authors).  We can cite, e.g., the role of the Planck constant $\hbar$ (which has to be treated as the central element in the Heisenberg algebra and as a constant operator \cite{cct}),
the introduction of a new dynamical Lie algebra generated by primitive elements (composite operators which have to be regarded as Hopf algebra generators of their own, see \cite{ckt}), the breaking of the rotational invariance generated by the twist (the Lie-rotational algebra recovered
by twisted generators under twisted brackets \cite{{wess},{cknt},{cpt}} being only a formal construction which does not
correspond to a dynamical symmetry, \cite{ckt}), etc.. It is commonly stated that the devil is in the details. Then, the aim of this paper is to work out the ``devilish details" involved in a Drinfel'd twist quantization based on first principles. Even if we explicitly present our results just for noncommutative oscillators we get, as a bonus, a broad scheme to be applied to the first-quantization of a generic non-relativistic Hamiltonian system.\par
Concerning the applications to the noncommutative oscillators we are able not only to recover,
but also to extend the results previously 
obtained in the literature (see \cite{{gamboa1},{gamboa2}} and \cite{{kijanka},{scholtz}}) for a generic central potential. The full power of the Hopf algebra scheme is at work here. We illustrate
this feature by comparing our results with the ones obtained by the important references \cite{{kijanka},{scholtz}}. In these papers the single-particle
spectrum for the noncommutative oscillators has been derived.  Our own results for the single-particle spectrum coincide with their findings.
On the other hand, the existence of the coproduct allows us to unambiguosly fix the deformed
$2$-particle Hamiltonian (and also, due to associativity, the deformed multiparticle
 Hamiltonian).  In \cite{{gamboa1},{gamboa2},{kijanka},{scholtz}} the noncommutative Hamiltonian for the multiparticle sector was not discussed. Indeed, it is quite hard to figure out which would be the correct Hamiltonian without the guiding principle of the coproduct. On the other hand, at an operatorial level, the knowledge of (at least) a $2$-particle operator is essential to detect whether the theory is truly deformed or not.  The deformed $2$-particle Hamiltonian is not additive  since it is not given by the sum of two deformed single-particle Hamiltonians. It turns out that this is the key physical issue  expressing the twist-deformation or noncommutativity of the dynamical system. Indeed,
the single-particle spectrum can be reproduced by a suitably chosen undeformed Hamiltonian.  Therefore, we need to perform measurements involving at least a $2$-particle operator in order to detect whether we are dealing with a deformed system or an undeformed one
(an undeformed $2$-particle Hamiltonian is obviously additive).\par  
Concerning $2$-particle and multiparticle operators we have that, applied to a module, the deformed coproduct of the deformed Hamiltonian is unitarily equivalent to the undeformed coproduct of the deformed Hamiltonian. In this latter case the $2$-particle Hamiltonian is invariant under the operator $P$ which exchanges the two particles. For the deformed coproduct
of the deformed Hamiltonian  the invariance under the permutation group is realized  in terms
of the operator
$P'= UPU^{-1}$, obtained from $P$ via the unitary transformation $U$ induced by the twist.
The deformed theory is therefore symmetric under particle exchange. \par
In the literature several investigations were made concerning the statistics of a deformed theory.
In \cite{cghs} the case of a twist-deformed quantum system at a finite temperature was discussed.
In \cite{bmpv} the statistics was investigated for a different type of quantization prescription,
based on the Groenewold-Moyal plane. Contrary to \cite{bmpv}, our quantization framework requires an ordinary Hilbert space, with the noncommutativity encoded in the deformed generators. 
The statistics was also discussed for a different twist and in a relativistic setting in \cite{bjp} and \cite{bp}. For the ${\kappa}$-Poincar\'e theories the exchange-symmetry is not present and difficult
issues are involved in trying to restore it, see \cite{aca}--\cite{luk}.
\par
For what concerns the rotational invariance of the oscillators under abelian Drinfel'd twist deformation the situation is as follows.
The two-dimensional deformed oscillator maintains the $so(2)$ rotational invariance
(this is essentially due to the existence of the $\epsilon_{12}$ constant antisymmetric tensor), while for the three-dimensional deformed oscillator the $so(3)$ rotational invariance is
broken down to an $so(2)$-invariance.\par 
The scheme of the paper is as follows. In Section {\bf 2} we introduce the main ingredients of our construction concerning the realization of a twist-deformed Hopf algebra on a Hilbert space. 
In particular we will introduce the dynamical Lie algebra of primitive elements (induced by a Heisenberg algebra) associated to the harmonic oscillators and discuss the framework of the
so-called ``hybrid quantization". In Section {\bf 3} we apply the previous formalism to the
quantization of the $d=2$ noncommutative oscillator. This Section is divided in three subsections presenting, respectively, the single-particle spectrum of the deformed Hamiltonian,
the construction of the multiparticle deformed Hamiltonian and the analysis of the rotational invariance in the deformed case. We will point out the unitary equivalence, when applied to a Hilbert space, of the deformed coproduct of the deformed Hamiltonian and the undeformed coproduct of the deformed Hamiltonian. This equivalence brings, as a consequence, the symmetry of the deformed multi-particle Hamiltonian under particle-exchange. In Section {\bf 4}
we repeat the previous Section steps to the case of the $d=3$ noncommutative oscillators.
We leave to the Conclusions a detailed discussion of our results and of their physical applications.
The paper is further complemented by two Appendices. In Appendix {\bf A}  we point out that
the non-additivity of the deformed $2$-particle Hamiltonian is essential to physically detect a deformation (a measurement based on single-particle operators cannot physically discriminate a constant non-commutativity from the undeformed case). In Appendix {\bf B} the most relevant formulas for the Drinfel'd twist deformation of a Hopf algebra are recollected.    
      
\section{The dynamical Lie algebra and its deformation}

The starting point is the $d$-dimensional Heisenberg algebra ${\cal H}_d$, regarded as a Lie algebra, with generators
$\hbar$ (a central charge), $x_i$ and $p_i$ ($i=1,2,\ldots, d$) satisfying
the commutation relations
\begin{eqnarray}\label{heis}
[x_ip_j]=i\hbar \delta_{ij}, &&  [\hbar,x_i]=[\hbar,p_i]=0.
\end{eqnarray}
It allows us to introduce the enlarged Lie algebra ${\cal G}_d$, containing ${\cal H}_d$ as
a subalgebra, together with the extra generators $H$, $K$, $D$ and $L_{i_1\ldots i_{d-2}}$:
\begin{equation}
 {\cal{G}}_d = \{ \hbar, x_i, p_i,  H, K, D, L_{i_1 i_2 \cdots i_{d-2}} \}, \; i = 1,...,d.
\end{equation}
The commutation relations among the ${\cal G}_d$ generators are recovered from the Heisenberg algebra relations, together with the identifications
\begin{eqnarray}{\label{ident}}
 H &=& \frac{1}{2\hbar} \left(p_ip_i \right), \nonumber\\ 
K &=& \frac{1}{2\hbar} \left(x_ix_i \right), \nonumber\\
 D&=& \frac{1}{4\hbar} \left(x_ip_i + p_ix_i \right),\nonumber\\ 
 L_{i_1 i_2 \cdots i_{d-2}} &=& \frac{1}{\hbar} \epsilon_{i_1 i_2 \cdots i_{d-1} i_d} x_{i_{d-1}} p_{i_d}
\end{eqnarray}
(the sum over repeated indices is understood).
\par
For $d=3$ we get, explicitly,
\begin{eqnarray}{\label{Gd}}
\relax [x_i,p_j]&=&i\hbar \delta_{ij},\nonumber\\
\relax [D,H ] &= & iH,\nonumber\\
\relax [D,K] &=& -iK, \nonumber\\
\relax [K,H] &=& 2iD,\nonumber \\
\relax [x_i,H] &=& ip_i,\nonumber\\ 
\relax [x_i,D] &=& \frac{i}{2}x_i,\nonumber \\ 
\relax [p_i,K] &=& -ix_i,\nonumber\\ 
\relax [p_i,D] &=& -\frac{i}{2}p_i,\nonumber\\ 
\relax [ L_i,  x_j] &=& i \epsilon_{i j k} x_k,\nonumber \\
\relax [L_i, p_j] &=& i \epsilon_{i j k} p_k ,\nonumber \\
\relax [ L_i, L_j] &=& i\epsilon_{ijk}L_k 
\end{eqnarray}
(the remaining commutation relations are vanishing).\par
For $d=2$ we have a single rotation generator $L$ such that 
\begin{eqnarray}
\relax [L, x_i]&=& i\epsilon_{ij}x_j,\nonumber\\
\relax [L,p_i]&=& i\epsilon_{ij} p_j
\end{eqnarray}
and vanishing commutation relations otherwise.\par
One should note that the presence of $\hbar$ in the denominator of the right hand side of (\ref{ident}) is required in order to define (\ref{Gd}) as a Lie algebra (${\cal G}_d$ is obtained from (\ref{heis}) and (\ref{ident}) as an abstract Poisson brackets algebra). The hamiltonian ${\bf H}$ of a $d$-dimensional harmonic oscillator is a linear combination of $H$ and $K$. We can therefore regard ${\cal G}_d$ as the dynamical Lie algebra of the harmonic oscillator containing, in particular,  the generators  $L_{i_1\cdots i_{d-2}}$ of the $d$-dimensional rotations.  
\par
One is now in position to introduce the Universal Enveloping Lie Algebra ${\cal U}({\cal G}_d)$ and endow it with a Hopf algebra structure (see the Appendix {\bf B} for a short review on Hopf algebras). In the undeformed case, the undeformed coproduct implies the additivity of the ${\cal G}_d$ generators
(therefore, in particular, the additivity of the momenta and of the energy). In \cite{ckt} the physical considerations leading to the introduction of a dynamical Lie algebra of ``primitive elements" (in the present case, the ${\cal G}_d$ generators) have been discussed at length and will not be repeated here. We limit ourselves to mention that the (\ref{ident}) identifications only hold at a Lie algebra level, but not as a Hopf algebra relation. From now on we will denote these weak identifications with the symbol ``$\approx$". For instance, in $d=3$, we get for the third component $L_z$ of the angular momentum the weak identification
\begin{eqnarray}
\hbar L_z &\approx & x p_y-yp_x.
\end{eqnarray}
 Indeed, the coproduct of the left hand side does not coincide with the coproduct of the right hand side. Similarly, the $\hbar$ generator can be identified with the ${\bf 1}$ identity operator only weakly
\begin{eqnarray}\label{weakh}
\hbar &\approx& {\bf 1}.
\end{eqnarray}
When representing the ${\cal G}_d$ generators as operators acting on a module we are dealing with the ``$\approx$" weak equivalence.

\subsection{The abelian twist-deformed ${\cal U}^{\cal F}({\cal G}_d)$ Hopf algebra}

The ${\cal U}({\cal G}_d)$ algebra can be deformed (see Appendix {\bf B}) via the twist 
\begin{eqnarray}\label{abtwist}
\cal{F} &=& \exp\left(i \alpha_{ij} p_i\otimes p_j\right), \quad \alpha_{ij}=-\alpha_{ji},
\end{eqnarray}
which is abelian since  $[p_i,p_j]=0$.  The twist is well-defined due to the fact that the $p_i$ momenta are among the generators of ${\cal G}_d$.
\par
The twist induces a deformation ($g\mapsto g^{\cal F}$) for the ${\cal G}_d$ generators, with $g^{\cal F}$ belonging to ${\cal U}({\cal G}_d)$. 
The generators which commute with $p_i$ remain undeformed and the same is true of $D$, due to a contraction between a symmetric and a skew-symmetric tensor. The deformed generators are
\begin{eqnarray}
 x_i^{\mathcal{F}} &=& x_i - \alpha_{ij} p_j \hbar, \nonumber \\
 K^{\mathcal{F}} &=& K - \alpha_{ij}x_ip_j + \frac{\alpha_{jk}\alpha_{jl}}{2!}p_kp_l \hbar, \nonumber \\
 L_{i_1 i_2 \cdots i_{d-2}}^{\cal{F}} &=&  L_{i_1 i_2 \cdots i_{d-2}} - \epsilon_{i_1 i_2 \cdots i_{d-2} j k}  \alpha_{j l}  p_k p_l .
\end{eqnarray} 
The deformation of the position operators $x_i$ corresponds to the Bopp shift.\par
For what concerns the brackets structure that we have to quantize, there are different possibilities which have been discussed in \cite{cct}.
One can use e.g.\ the twist-deformed brackets $[\cdot,\cdot]_{\cal F}$, see Appendix {\bf B}, or the original commutators. The latter choice is the most convenient in application to quantization since the brackets between two operators are given by ordinary commutators. The presentation of the ${\cal U}^{\cal F}({\cal G}_d)$ deformed Hopf algebra in terms of deformed generators and ordinary commutators is known as the ``hybrid" formalism \cite{cct}. It is in connection with the hybrid formalism that we succeed to link the abelian twist to the (constant) noncommutativity. Indeed,
\begin{eqnarray}
[x_i^{\cal F}, x_j^{\cal F}]&=& i\Theta_{ij},
\end{eqnarray}
where the constant operator $\Theta_{ij}$ is given by
\begin{eqnarray}
\Theta_{ij}&=&2\alpha_{ij}\hbar^2.
\end{eqnarray}
The final justification for using the hybrid formalism is the fact that it produces a self-consistent non-trivial deformation in the multi-particle sector.
 For single-particle operators the knowledge of the deformed generators, together with their commutators and their action on a module $V$ which possesses the structure of a Hilbert space,
is sufficient to quantize the system. For multi-particle operators the extra-structure of the (deformed) coproduct plays a role. The deformed $2$-particle operator associated with the
deformed generator $g^{\cal F}$ is constructed by applying $\Delta^{\cal F}(g^{\cal F})\in 
{\cal U}^{\cal F}({\cal G}_d)\otimes  {\cal U}^{\cal F}({\cal G}_d)$ to the Hilbert space $V\otimes V$. The twist ${\cal F}$ (\ref{abtwist}), applied to $V\otimes V$, corresponds to the unitary operator $F$.
Since 
\begin{eqnarray}
\Delta^{\cal F}(g^{\cal F}) &=& {\cal F}\cdot \Delta (g^{\cal F})\cdot  {\cal F}^{-1},\nonumber\\
\end{eqnarray}
with $\Delta (g^{\cal F})$ the undeformed coproduct, we end up that the operators
$\widehat{\Delta^{\cal F}}(g^{\cal F})$, $\widehat{\Delta}(g^{\cal F})$, acting on $V\otimes V$, are unitarily equivalent:
\begin{eqnarray}
\widehat{\Delta^{\cal F}}(g^{\cal F}) &=& F\cdot \widehat{\Delta}(g^{\cal F})\cdot F^{-1}.
\end{eqnarray}
This feature also applies for $n$-particle operators with $n\geq 3$. \par
It is convenient to introduce the symbol `` ${\widehat ~}$ " when we need to make the distinction
between an element $\Omega$ of the (tensor product of the) Universal Enveloping Lie Algebra and
its action ${\widehat \Omega}$ on a module. Therefore, ${\bf H}^{\cal F}\in {\cal U}^{\cal F}({\cal G}_d)$ while ${\widehat {\bf H}^{\cal F}}:V\rightarrow V$.\par
We have at this point a viable scheme to first-quantize an abelian twist-deformed (noncommutative) quantum mechanical system based on the following steps. We introduce at first the dynamical Lie algebra ${\cal G}_d$. Next, we represent it on the Hilbert space $V$.
Later we introduce the twist-deformation and realize the single-particle operators as deformed generators $g^{\cal F}$ acting on $V$. The multi-particle operators are constructed by applying
the (undeformed) coproducts of $g^{\cal F}$ on the tensor space $V\otimes \ldots \otimes V$.  
\par
In the following we will show that the choice of using the undeformed coproduct (instead of the unitarily equivalent deformed coproduct) is particularly useful since $\Delta ({\bf H}^{\cal F})$,
 i.e. the deformed $2$-particle Hamiltonian, turns out to be automatically symmetric in the exchange between first and second particle.   

\section{The two-dimensional twisted oscillator}

In $d=2$ we can express the deformation parameter $\alpha_{ij}$ as
\begin{eqnarray}
\alpha_{12}&=&\epsilon_{12}\frac{\alpha}{Z}.
\end{eqnarray}
From Eq. (\ref{abtwist}) we have that $\alpha_{12}$ is a dimensional parameter. $Z$ is a constant unit reference with the dimension of $[p^2]$, the square of the momentum,
so that $\alpha$ is a non-dimensional parameter. $Z$ can be expressed in terms of the mass and the energy separation between adjacent eigenvalues of the undeformed harmonic oscillator. In our conventions we choose to set $Z=1$. 
Without loss of generality we can restrict $\alpha$ to belong to the fundamental domain
$\alpha\in [0,+\infty]$. The value $\alpha=0$ corresponds to the undeformed $2$-dimensional harmonic oscillator, while the $\alpha\rightarrow+\infty $ limit is non-singular as shown later.\par
For the construction of the Hilbert space $V$ of the $2$-dimensional harmonic oscillator with
Hamiltonian ${\bf H}=H+K$ (see (\ref{ident})) we introduce the creation and annihilation operators $a_i$ and $a_i^\dagger$, such that
\begin{eqnarray}\label{cre}
a_i&=&\frac{x_i - ip_i}{\sqrt{2}}, \nonumber\\
a_i^\dagger&=&\frac{x_i + ip_i}{\sqrt{2}},
\end{eqnarray} 
where
\begin{eqnarray}
[a_i,a_j^\dagger]&=&\hbar\delta_{ij}
\end{eqnarray}
for $i,j=1,2$. \par
A different basis, given by $b_\pm$, $b_\pm^\dagger$, is more convenient for constructing the deformed spectrum:
\begin{eqnarray}
b_\pm&=&\frac{a_x\mp ia_y}{\sqrt{2}},\nonumber \\
b_\pm^\dagger &=&\frac{a_x^\dagger\pm ia_y^\dagger}{\sqrt{2}},
\end{eqnarray}
with
\begin{eqnarray}
[b_\pm,b_\pm^\dagger]&=&\hbar.
\end{eqnarray}
It is such that
\begin{eqnarray}
\relax[{\bf H},b_\pm ]&=&-b_\pm,\nonumber\\
\relax[{\bf H},b_\pm^\dagger ]&=&b_\pm^\dagger.
\end{eqnarray}
In $d=2$ the angular momentum $L$ is a scalar. It satisfies the commutation relations
\begin{eqnarray}
\relax[L,b_\pm]&=& \mp b_\pm,\nonumber\\
\relax [L,b_\pm^\dagger]&=& \pm b_\pm^\dagger.
\end{eqnarray}
The deformed Hamiltonian 
${\bf H}^{\cal F}\in {\cal U}^{\cal F}({\cal G}_2)$ 
is given by the expression
\begin{eqnarray}
\bf{H}^{\cal F}&=&H^{\cal F}+K^{\cal F}=H+K -\alpha xp_y+\alpha yp_x+\frac{\alpha^2}{2}\hbar (p^2_x + p^2_y).
\end{eqnarray}
The deformed $2$-particle Hamiltonian 
belonging to $ {\cal U}^{\cal F}({\cal G}_2)\otimes  {\cal U}^{\cal F}({\cal G}_2)$
is unambiguously fixed in terms of the coproduct,
as discussed in Section {\bf 2}. We have
\begin{eqnarray}
\Delta({\bf H}^{\cal F})&=&{\bf H}^{\cal F}\otimes{\bf 1}+{\bf 1}\otimes{\bf H}^{\cal F}+\alpha(y\otimes p_x+p_x\otimes y
-x\otimes p_y-p_y\otimes x) \nonumber \\
&&+\frac{\alpha^2}{2} \sum_{i=1}^2 (2p_i\hbar\otimes p_i+2p_i\otimes p_i\hbar+p_i^2\otimes\hbar+\hbar\otimes p_i^2).
\end{eqnarray}
One should note that the right hand side is symmetric in the exchange of the first with the second particle.

\subsection{The single-particle spectrum}

Throughout this subsection we apply the deformed operators to the module, given by the Fock space $V$ obtained by repeatedly applying the creation operators $b_{\pm}^\dagger$ on the Fock vacuum $|0\rangle$ ($b_\pm|0\rangle =0$).  We are entitled to weakly set $\hbar=1$,
see Eq. (\ref{weakh}).  Since we are dealing with operators acting on a module, all equalities in this subsection have to be understood as weak equalities. 
In order not to make our notation unnecessarily heavy we avoid here using the `` ${\widehat ~}$ " symbol for operators. No confusion will arise. 
\par
In terms of the $b_\pm,b_\pm^\dagger$ operators the Hamiltonian reads as
\begin{eqnarray}
\bf{H}&=&\frac{1}{2}\sum_{i=\pm}\{b_i,b^\dagger_i\},
\end{eqnarray}
while a number operator $N$ and the angular momentum $L$ can be expressed as
\begin{eqnarray}
N&=&b^\dagger_+b_++b^\dagger_-b_-=N_++N_-,\nonumber \\
L&=&b^\dagger_+b_+-b^\dagger_-b_-=N_+-N_-.
\end{eqnarray}
Since $[{\bf H},L]=0$, the $|n_+ n_-\rangle$ basis simultaneously diagonalizes  both operators:
\begin{eqnarray}
{\bf H}|n_+ n_-\rangle&=&(n_++n_-+1)|n_+ n_-\rangle,\nonumber\\
L|n_+ n_-\rangle&=&(n_+-n_-)|n_+ n_-\rangle.
\end{eqnarray}
It is convenient to reexpress the above results through the integers $n=n_++n_-$ and 
$m=n_+-n_-$, so that
\begin{eqnarray}
\mathbf{H}|nm\rangle&=&(n+1)|nm\rangle,\nonumber\\
L|nm\rangle&=&m|nm\rangle.
\end{eqnarray}
The deformed Hamiltonian
\begin{eqnarray}
{\bf H}^{\cal F}&=&H^{\cal F}+K^{\cal F}=H+K-\alpha xp_y+\alpha yp_x+\frac{\alpha^2}{2}(p^2_x + p^2_y),
\end{eqnarray}
applied to the Hilbert space $V$, can be reproduced by the linear combination
\begin{eqnarray}\label{unren2}
{\bf H}^{\cal F}&=&{\widetilde{\bf H}}-\alpha L,
\end{eqnarray}
where
\begin{eqnarray}
{\widetilde{\bf{H}}}&=&(1+\alpha^2)H+K
\end{eqnarray}
can be regarded as a renormalized undeformed Hamiltonian. \par
In terms of the above expressions the spectrum of ${\bf H}^{\cal F}$ can be easily computed:
\begin{eqnarray}
{\bf H}^{\cal F}|nm\rangle&=&({\widetilde{\bf{H}}}-\alpha L)|nm\rangle=\left[(\sqrt{1+\alpha^2})(n+1)-\alpha m\right]|nm\rangle,
\end{eqnarray}
for  $m=-n,-n+2,\ldots,n-2,n$ and $n$ a non-negative integer.
\par
The vacuum is recovered for $n=0$ (the vacuum energy is $\sqrt{1+\alpha^2}$).
\par
For $n=1,2$ we have
\begin{eqnarray}
|1,1\rangle&:&2\sqrt{1+\alpha^2}-\alpha,\nonumber\\
|1,-1\rangle&:&2\sqrt{1+\alpha^2}+\alpha,\nonumber\\
|2,2\rangle&:&3\sqrt{1+\alpha^2}-2\alpha,\nonumber\\
|2,0\rangle&:&3\sqrt{1+\alpha^2},\nonumber\\
|2,-2\rangle&:&3\sqrt{1+\alpha^2}+2\alpha.
\end{eqnarray}
This spectrum coincides with the one computed in \cite{{kijanka},{scholtz}}.\par
It is worth pointing out that, in the limit for $\alpha\rightarrow +\infty$, the normalized Hamiltonian 
\begin{eqnarray}
{\bf H}_N^{\cal F} &=& \frac{1}{\sqrt{1+\alpha^2}}{\bf H}^{\cal F}
\end{eqnarray}
is well-defined and coincides with the identity operator ${\bf 1}$ acting on the reduced Hilbert
space $V'\subset V$ spanned by the vectors $|n,n\rangle$. The remaining vectors decouple from the theory because their energy gap
with respect to the degenerate vacuum tends to infinity.

\subsection{The multi-particle Hamiltonian}

The deformed two-particle Hamiltonian acting on $V\otimes V$ is recovered by computing, at first, the (undeformed) coproduct of the deformed Hamiltonian ${\bf H}^{\cal F}$.  We have 
\begin{eqnarray}\label{deltadef}
\Delta(\mathbf{H}^\mathcal{F})&=&
\mathbf{H}^\mathcal{F}\otimes\mathbf{1}+\mathbf{1}\otimes\mathbf{H}^\mathcal{F}+\alpha(y\otimes p_x+p_x\otimes y
-x\otimes p_y-p_y\otimes x) \nonumber \\
&&+\frac{\alpha^2}{2}\sum_{i=1}^2(2p_i\hbar\otimes p_i+2p_i\otimes p_i\hbar+p_i^2\otimes\hbar+\hbar\otimes p_i^2).
\end{eqnarray}
As discussed in the previous Section, the convenience in using the undeformed coproduct
(unitarily equivalent to the use of the deformed coproduct, when applied to the $V\otimes V$ Hilbert space)
lies on the fact that the right hand side is in this case automatically symmetric in the particle
exchange. One should also note that the deformed two-particle Hamiltonian is no longer additive due to the extra terms dependent on $\alpha$.  Even if no longer additive, the coassociativity of the coproduct guarantees in any case the associativity of the deformed Hamiltonian. Indeed, for three-particle states, we have the equality  
\begin{equation}
(id\otimes\Delta)\Delta(\mathbf{H}^\mathcal{F})=(\Delta\otimes id)\Delta(\mathbf{H}^\mathcal{F})\equiv\Delta_{(2)}(\mathbf{H}^\mathcal{F}),
\end{equation}
where, explicitly 
\begin{eqnarray}\label{delta2}
\Delta_{(2)}(\mathbf{H}^\mathcal{F})&=&\mathbf{H}^\mathcal{F}\otimes
\mathbf{1}\otimes\mathbf{1}+\mathbf{1}\otimes\mathbf{H}^\mathcal{F}
\otimes\mathbf{1}+\mathbf{1}\otimes\mathbf{1}\otimes\mathbf{H}^\mathcal{F} \nonumber\\
&&+\alpha(\mathbf{1}\otimes y\otimes p_x+y\otimes\mathbf{1}\otimes p_x+y\otimes p_x\otimes\mathbf{1})\nonumber\\
&&+\alpha(\mathbf{1}\otimes p_x\otimes y+p_x\otimes\mathbf{1}\otimes y+p_x\otimes y\otimes\mathbf{1})\nonumber\\
&&-\alpha(\mathbf{1}\otimes x\otimes p_y+x\otimes\mathbf{1}\otimes p_y+x\otimes p_y\otimes\mathbf{1})\nonumber\\
&&-\alpha(\mathbf{1}\otimes p_y\otimes x+p_y\otimes\mathbf{1}\otimes x+p_y\otimes x\otimes\mathbf{1})\nonumber\\
&&+\alpha^2\sum_{i=1}^2[\mathbf{1}\otimes p_i\hbar\otimes p_i+p_i\hbar\otimes p_i\otimes\mathbf{1}+p_i\hbar\otimes p_i\otimes\mathbf{1}\nonumber\\
&&+\mathbf{1}\otimes p_i\otimes p_i\hbar+p_i\otimes p_i\hbar\otimes\mathbf{1}+p_i\otimes p_i\hbar\otimes\mathbf{1}\nonumber\\
&&+\hbar\otimes p_i\otimes p_i+p_i\otimes p_i\otimes\hbar+p_i\otimes p_i\otimes\hbar\nonumber\\
&&+\frac{1}{2}(\mathbf{1}\otimes \hbar\otimes p_i^2+\hbar\otimes p_i^2\otimes\mathbf{1}+\hbar\otimes p_i^2\otimes\mathbf{1}\nonumber\\
&&+\mathbf{1}\otimes p_i^2\otimes \hbar+p_i^2\otimes \hbar\otimes\mathbf{1}+p_i^2\otimes \hbar\otimes\mathbf{1})].
\end{eqnarray}
The deformed two-particle energy $E_{12}^{\cal F}$ can be expressed as
\begin{eqnarray}\label{defaddit}
E_{12}^{\cal F} &=& E_1^{\cal F}+E_2^{\cal F}+\Omega_{12},
\end{eqnarray}
where $E_i^{\cal F}$ ($i=1,2$) are the single-particle energies and $\Omega_{12}$ is an effective interaction term (we avoid calling it an interacting potential because it depends on the momenta as well as on the position operators). Therefore we have at least two possible interpretations for the above results. Either we regard $\Omega_{12}$ as an interaction or we regard (\ref{defaddit}) as describing a system of free (albeit deformed) particles, with $\Omega_{12}\neq 0$ as a measure of deformation.\par
The associativity is expressed by the three-particle formula
\begin{eqnarray}\label{intter2}
E_{123}^{\cal F} &\equiv & E_{(12)3}^{\cal F}=E_{1(23)}^{\cal F}=E_{1}^{\cal F}+E_{2}^{\cal F}+E_{3}^{\cal F}+\Omega_{12}+\Omega_{23}+\Omega_{31} +\Omega_{123},
\end{eqnarray}
with $\Omega_{123}$ recovered from the $\Omega_{ij}$'s.\par
It should be stressed the crucial role of the coproduct in unambiguously determine the ``interacting term" $\Omega_{12}$.\par
The formulas (\ref{deltadef}) and (\ref{delta2}) are equalities in the tensor products of the Universal Enveloping Lie algebras ${\cal U}^{\cal F}({\cal G}_2)\otimes {\cal U}^{\cal F}({\cal G}_2)$
and ${\cal U}^{\cal F}({\cal G}_2)\otimes {\cal U}^{\cal F}({\cal G}_2)\otimes {\cal U}^{\cal F}({\cal G}_2)$, respectively.
We specialize them now as operator equalities acting on the $V\otimes \ldots\otimes V$ 
multi-particle Hilbert space. As discussed before $\hbar$ is mapped into the identity operator (\ref{weakh}). 
By expressing $\Omega_1 = {\widehat \Omega} \otimes {\bf 1}$ and
$\Omega_2 = {\bf 1}\otimes {\widehat \Omega}$,
we can write the deformed two-particle Hamiltonian operator  ${\widehat \Delta} (\mathbf{H}^{\mathcal{F}})$ acting on $V\otimes V$ as
\begin{equation}
 {\widehat\Delta} (\mathbf{H}^{\mathcal{F}}) = {\mathbf{H}}^{\mathcal{F}}_{1} + {\mathbf{H}}^{\mathcal{F}}_{2} + 
\alpha (\epsilon_{ij}p^{(1)}_ix^{(2)}_j - \epsilon_{ij}x^{(1)}_ip^{(2)}_j) + \alpha^2\left(
\frac{1}{2}  ( p^{(1)}_ip^{(1)}_i + p^{(2)}_ip^{(2)}_i ) +
2  p^{(1)}_ip^{(2)}_i\right).
\end{equation}
We can write
\begin{eqnarray}
 {\widehat  \Delta} (\mathbf{H}^{\mathcal{F}}) &=& {\mathbf{H}}^{\mathcal{F}}_{1} + {\mathbf{H}}^{\mathcal{F}}_{2} +{\widehat \Omega}_{12},
\end{eqnarray}
where the interacting term ${\widehat \Omega}_{12}$ can be written, in terms of the creation and annihilation operators, as
\begin{eqnarray}\label{omega2}
{\widehat \Omega}_{12}&=&
\alpha(1+ \alpha)(b^{(1)}_+ b^{\dagger \; (2)}_+ + 
b^{(2)}_+ b^{\dagger \; (1)}_+)- \alpha(1- \alpha)(b^{(1)}_- b^{\dagger \; (2)}_- + b^{(2)}_- b^{\dagger \; (1)}_-)\nonumber\\
&&
+  \alpha^2(b^{(1)}_+ b^{(2)}_- + b^{(1)}_- b^{(2)}_+ + 
b^{\dagger \; (1)}_+ b^{\dagger \; (2)}_- + b^{\dagger \; (1)}_- b^{\dagger \; (2)}_+).
\end{eqnarray}
The computation of the interacting term ${\widehat \Omega}_{12}$ goes beyond the results of
\cite{{kijanka},{scholtz}}.

\subsection{The rotational invariance}

The  undeformed harmonic oscillator is rotationally invariant. In $d=2$ the deformed oscillator maintains the $so(2)$ rotational  invariance \cite{ckt}. This result can be recovered as follows. 
The $L$ generator of the rotation on the $xy$ plane, deformed under the twist according to
\begin{equation}
 L^{\mathcal{F}} = L - \alpha (p_x^2+p_y^2),
\end{equation}
 is no longer a rotation generator in the deformed case since
\begin{equation}
 [L^{\mathcal{F}},x_i^{\mathcal F}] = i (\epsilon_{ij}x_j^{\mathcal F} + 2 \alpha  p_i \hbar).
\end{equation}
Furthermore, $L^{\cal F}$  does not commute with the deformed hamiltonian ${\bf H}^{\cal F}$.\par
 On the other hand, there exists in the ${\cal U}^{\cal F}({\cal G}_2)$ Universal Enveloping Algebra a generator possessing the above properties. It is the original generator $L$ itself. We have indeed that
\begin{eqnarray}
[L, {x_i}^{\cal F}]&=&i\epsilon_{ij}x_j^{\cal F}
\end{eqnarray}
and
\begin{eqnarray}
[L, {\bf H}^{\cal F}]&=&0.
\end{eqnarray}
We are therefore assured that the single-particle spectrum is rotationally invariant. The rotational invariance is also maintained in the multi-particle sector due to the commutation
relation
\begin{equation}
 \left[ \Delta(\mathbf{H}^{\mathcal{F}}), \Delta(L)\right] = 0.
\end{equation}

\section{The three-dimensional twisted oscillator}

We can repeat for $d=3$ the same steps that we discussed for the $d=2$ harmonic oscillator.
The deformation parameter $\alpha_{ij}$ can now be expressed as
\begin{eqnarray}
\alpha_{ij}&=&\epsilon_{ijk}\frac{\alpha_k}{Z}.
\end{eqnarray}
As before we set the dimensional reference unit $Z=1$.  We are therefore left with the non-dimensional vector ${\vec \alpha}=(\alpha_1,\alpha_2,\alpha_3)$. Without loss of generality
we can choose the $xyz$ coordinate system in such a way that the third direction points towards
the ${\vec\alpha}$ direction. We can therefore set ${\vec\alpha}=(0,0,\alpha_3=\alpha)$. As before, the fundamental domain for $\alpha$ is $\alpha\in [0,+\infty ]$.\par
Without loss of generality we can define the undeformed $d=3$ Hamiltonian of the harmonic
oscillator as ${\bf H}=H+K$ (where $H,K\in\mathcal{G}_3$). 

To construct the Hilbert space $V$ of the three-dimensional harmonic oscillator, we shall proceed similarly to the $d=2$ case. We initially introduce the same creation and annihilation operators of (\ref{cre}), but now with $i=1,2,3$, and similarly perform the change of basis
\begin{eqnarray}
b_\pm&=&\frac{a_x\mp ia_y}{\sqrt{2}},\nonumber \\
b_\pm^\dagger &=&\frac{a_x^\dagger\pm ia_y^\dagger}{\sqrt{2}},\nonumber\\
b_z&=&a_z,\nonumber\\
b_z^\dagger&=&a_z^\dagger,
\end{eqnarray}
satisfying, as expected for creation and annihilation operators,
\begin{eqnarray}
[b_i,b_j^\dagger]&=&\delta_{ij}\hbar,\nonumber\\
\relax[{\bf H},b_i ]&=&-b_i,\nonumber\\
\relax[{\bf H},b_i^\dagger ]&=&b_i^\dagger.
\end{eqnarray}
with $i=\pm,z$.

Once applied the abelian twist-deformation (\ref{abtwist}), we have that the deformed Hamiltonian
reads as
\begin{equation}
\mathbf{H}^\mathcal{F} = H+K-\alpha (xp_y- yp_x)+\frac{\alpha^2}{2}\hbar(p_x^2+p_y^2),
\end{equation}
and the undeformed coproduct of the deformed Hamiltonian is formally similar to the $d=2$ case:
\begin{eqnarray}\label{defcop3}
\Delta(\mathbf{H}^\mathcal{F})&=&\mathbf{H}^\mathcal{F}\otimes\mathbf{1}+\mathbf{1}\otimes\mathbf{H}^\mathcal{F}+\alpha(y\otimes p_x+p_x\otimes y
-x\otimes p_y-p_y\otimes x) \nonumber \\
&+&\frac{\alpha^2}{2}\sum_{i=1}^2(2p_i\hbar\otimes p_i+2p_i\otimes p_i\hbar+p_i^2\otimes\hbar+\hbar\otimes p_i^2),
\end{eqnarray}
which is symmetric under particle exchange.

In $d=3$ the angular momentum is a vector given in (\ref{Gd}).

\subsection{The single-particle spectrum}

The deformed single-particle spectrum is obtained by applying the deformed operators to
the Fock space $V$, constructed in terms of the $d=3$ creation operators acting on the $d=3$ Fock vacuum. We use the same conventions as in the $d=2$ computations (we set $\hbar=1$, we drop
the `` ${\widehat ~}$" symbol for operators, etc.).

Let us first express the Hamiltonian as
\begin{eqnarray}
\bf{H}&=&\frac{1}{2}\sum_{i=\pm,z}\{b_i,b^\dagger_i\},
\end{eqnarray}
and recall that in this basis one can write the operators
\begin{eqnarray}
N_{xy}&=&b^\dagger_+b_++b^\dagger_-b_-=N_++N_-,\nonumber \\
N_z&=&b_z^\dagger b_z,\nonumber \\
L_z&=&b^\dagger_+b_+-b^\dagger_-b_-=N_+-N_-.
\end{eqnarray}

The spectrum of the undeformed Hamiltonian is given in terms of the three non-negative
integers $n_\pm, n_z$ as
\begin{equation}
{\bf H}|n_+ n_- n_z\rangle=\left(n_++n_-+n_z+\frac{3}{2}\right)|n_+ n_-n_z\rangle,
\end{equation}
while the action of the third component of the angular momentum reads
\begin{equation}
L_z|n_+ n_-n_z\rangle=(n_+-n_-)|n_+ n_-n_z\rangle.
\end{equation}

It is convenient to relabel the states by means of the integers $n_{xy}=n_++n_-$ and 
$m=n_+-n_-$, and thus
\begin{eqnarray}
\mathbf{H}|n_{xy}n_zm\rangle&=&\left(n_{xy}+n_z+\frac{3}{2}\right)|n_{xy}n_zm\rangle,\nonumber\\
L_z|nm\rangle&=&m|n_{xy}n_zm\rangle.
\end{eqnarray}

It is now crucial to perform a splitting in the Hamiltonian. At the algebric level, the weak equality
\begin{equation}
\mathbf{H}=\frac{1}{2}\sum_{i=1}^3(x_i^2+p_i^2)
\end{equation}
holds. We can therefore separate $\mathbf{H}$ into its $xy$-part and its $z$-part:
\begin{equation}
\mathbf{H}=\mathbf{H}_{xy}+\mathbf{H}_z,
\end{equation}
where $\mathbf{H}_{xy}=\frac{1}{2}(x^2+p_x^2+y^2+p_y^2)$ and $\mathbf{H}_z=\frac{1}{2}(z^2+p_z^2)$.

We are now able to calculate the spectrum of the deformed Hamiltonian. It is clear that the deformation 
only affects $\mathbf{H}_{xy}$ and that the deformed Hamiltonian
\begin{equation}
\mathbf{H}^\mathcal{F} = H+K-\alpha (xp_y- yp_x)+\frac{\alpha^2}{2}(p_x^2+p_y^2)
\end{equation}
can be written as
\begin{equation}\label{unren3}
\mathbf{H}^\mathcal{F}=\widetilde{\mathbf{H}}_{xy}-\alpha L_z+\mathbf{H}_z,
\end{equation} 
where $\widetilde{\mathbf{H}}_{xy}$ is again a two-dimensional undeformed Hamiltonian with frequency $\tilde{\omega}=\sqrt{1+\alpha^2}$. It becomes evident that isotropy is lost.

Therefore, the spectrum of $\mathbf{H}^\mathcal{F}$ is
\begin{equation}
\mathbf{H}^\mathcal{F}|n_{xy}n_zm\rangle=\left[\sqrt{1+\alpha^2}(n_{xy}+1)-\alpha m+\left(n_z+\frac{1}{2}\right)\right]|n_{xy}n_zm\rangle,
\end{equation}
with $m=-n_{xy},-n_{xy}+2,\ldots,n_{xy}-2,n_{xy}$. 

Below we present a few states and their energies.

The vacuum is recovered for $n_{xy}=n_z=0$ and its energy is $\frac{1}{2}+\sqrt{1+\alpha^2}$.

For $n_{xy}+n_z=1$ we have
\begin{eqnarray}
|0,1,0\rangle&:&\frac{3}{2}+\sqrt{1+\alpha^2},\nonumber\\
|1,0,-1\rangle&:&\frac{1}{2}+2\sqrt{1+\alpha^2}+\alpha,\nonumber\\
|1,0,1\rangle&:&\frac{1}{2}+2\sqrt{1+\alpha^2}-\alpha .
\end{eqnarray}

For $n_{xy}+n_z=2$ we have
\begin{eqnarray}
|0,2,0\rangle&:&\frac{5}{2}+\sqrt{1+\alpha^2}, \nonumber\\
|1,1,-1\rangle&:&\frac{3}{2}+2\sqrt{1+\alpha^2}+\alpha,\nonumber\\
|1,1,1\rangle&:&\frac{3}{2}+2\sqrt{1+\alpha^2}-\alpha,\nonumber\\
|2,0,-2\rangle&:&\frac{1}{2}+3\sqrt{1+\alpha^2}+2\alpha,\nonumber\\
|2,0,0\rangle&:&\frac{1}{2}+3\sqrt{1+\alpha^2},\nonumber\\
|2,0,2\rangle&:&\frac{1}{2}+3\sqrt{1+\alpha^2}-2\alpha .
\end{eqnarray}

It can be noted that the energy of the states with $n_z=0$ coincides with the two-dimensional spectrum up to an additive factor of $\frac{1}{2}$, 
which is the zero-point energy along the $z$-axis.

\subsection{The multi-particle Hamiltonian}

Since the $\mathbf{H}_z$-part of the Hamiltonian remains untouched by the deformation, the ``interaction'' terms are formally exactly the same as in the two-dimensional case, with the difference that now $\mathbf{H}^\mathcal{F}$ is the three-dimensional deformed Hamiltonian belonging to $\mathcal{U}^{\mathcal{F}}(\mathcal{G}_3)$. 

In particular, for instance, the undeformed coproduct of the deformed Hamiltonian (\ref{defcop3}) is unitarily equivalent (acting on $V\otimes V$) to its deformed coproduct. It reads the same as (\ref{deltadef}), and so do the expressions of the energy of the three-particle state (\ref{delta2}), of the deformed two-particle energy (\ref{defaddit}) and of the energy associativity (\ref{intter2}).

\subsection{The rotational invariance}
The deformation of the angular momentum is given by the expression
\begin{equation}
 L_i^{\mathcal{F}} = L_i + \alpha p_i p_z-\alpha_i p_jp_j.
\end{equation}
Explicitly, this means that ($i=1,2,3$ is associated with the axis $x,y,z$, respectively)
\begin{eqnarray}
 L_x^{\mathcal{F}} &=& L_x + \alpha p_xp_z,\nonumber\\
 L_y^{\mathcal{F}} &=& L_y + \alpha p_yp_z,\nonumber\\
 L_z^{\mathcal{F}} &=& L_z - \alpha(p_x^2 + p_y^2).
\end{eqnarray}

They are not be generators of rotational symmetry, since
\begin{equation}
 [L_i^{\mathcal{F}},x_j^{\mathcal F}] = i \epsilon_{ijk}x_k^{\mathcal F} - 2 i \hbar (\delta_{ij}\alpha p_z-\alpha_ip_j).
\end{equation}

If we now perform the same computation with the $L_i$'s,
\begin{equation}
 [L_i,x_j^{\mathcal F}] = i \epsilon_{ijk}x_k^{\mathcal F} - i \hbar (\delta_{ij}\alpha p_z-p_i\alpha_j),
\end{equation}
we see that the second term of this expression vanishes only for $i=3$. Moreover, it is clear that $[\mathbf{H}^{\mathcal{F}},L_i]$ also vanishes only for $i=3$. We are therefore led to the conclusion that $L_z$ is a generator of rotational symmetry, while $L_x$ and $L_y$ are not. In the Universal Enveloping Lie Algebra there are no elements \cite{ckt} closing the $so(3)$ algebra and commuting with ${\bf H}^{\cal F}$.   This means that the original $so(3)$ rotational symmetry is broken down to a $so(2)$ symmetry by the deformation.

The same is true for the multi-particle states, since
\begin{eqnarray}
\relax [ \Delta(\mathbf{H}^{\mathcal{F}}), \Delta(L_z)] &=& 0.
\end{eqnarray}

\section{Conclusions}

In this paper we discussed a framework, based on a series of steps which cannot be bypassed,
to consistently quantize (in the first-quantization scheme) a twist-deformed dynamical system.
\par
Our scheme can be summarized as follows. At first we introduce the $d$-dimensional Heisenberg algebra ${\cal H}_d$ (as a Lie algebra, with $\hbar$ a central element).  In terms of the Heisenberg algebra generators we introduce a dynamical Lie algebra ${\cal G}_d$ containing the Hamiltonian operator, the momenta $p_i$'s, as well as other generators. For the harmonic oscillator the algebra ${\cal G}_d$ is finite-dimensional. For more general potentials, the algebra ${\cal G}_d$ can be infinite-dimensional (see \cite{ckt}). The (undeformed) Hopf algebra structure is introduced on the Universal Enveloping Algebra ${\cal U}({\cal G}_d)$. This is a crucial step. As discussed in \cite{cct} and \cite{ckt}, we are led to inconsistencies when trying to encode a physical system within the Hopf algebra structure of the Universal  Enveloping Algebra ${\cal U}({\cal H}_d)$, based on the Heisenberg algebra. Next, we apply the abelian Drinfel'd twist and construct the deformed Hopf algebra ${\cal U}^{\cal F}({\cal G}_d)$. The physical quantities are
associated with ${\cal F}$-deformed generators and their coproducts. 

The following step is introducing the Hilbert space $V$ as the module upon which ${\cal U}^{\cal F}({\cal G}_d)$ acts. We adopt the so-called hybrid formalism prescription which leads to a consistent quantization.
$V$ is the Fock space defined by the original undeformed creation and annihilation operators. We are therefore in the position to compute the deformed single-particle operators. We recover, for the noncommutative harmonic oscillators, the energy spectrum computed in \cite{kijanka, scholtz}. A key issue of our framework is that it allows us to further compute deformed multi-particle operators from the coproducts of the deformed generators.  Acting on $V\otimes \ldots\otimes V$, the undeformed coproduct is unitarily equivalent to the deformed coproduct. We argued,
for the systems that we investigated, the convenience of using the undeformed coproduct, because in this case the symmetry under particle exchange becomes manifest ($\Delta({\bf H}^{\cal F})$ is symmetric under particle exchange, while $\Delta^{\cal F}({\bf H}^{\cal F})$ is not). \par
The energy is no longer additive in the deformed case, but it maintains the associativity as a consequence of the coassociativity of the coproduct. On physical grounds, the most important byproduct of our method is the computation of the ${\widehat \Omega_{12}}$ term
entering either (\ref{defaddit}) or (\ref{intter2}).  ${\widehat \Omega_{12}}$ can be regarded as an effective interacting term, induced by the deformation,  between two free oscillators. \par

We pointed out that, at an operational level, the abelian twist-deformation cannot be detected by performing measurements on single-particle operators alone. Indeed, the spectrum of the
single-particle deformed Hamiltonian ${\bf H}^{\cal F}$ can be reproduced by a linear combination (see (\ref{unren2}) and (\ref{unren3})) of undeformed generators belonging to ${\cal G}_d$.
In the undeformed case the energy, however, is additive, so that in particular ${\widehat \Omega_{12}}=0$. Therefore, we cannot claim that our dynamical system is truly deformed unless we measure a multi-particle operator (at least a two-particle operator) and find out that ${\widehat \Omega_{12}}\neq 0$.\par
We used the presented scheme to first-quantize the non-commutative oscillators in presence of 
a constant non-commutativity. We were able to extend the results presented in the literature
by computing the multi-particle operators.
We furthermore checked out that the $so(2)$ rotational invariance of the $d=2$ oscillator is
preserved under deformation, while the $so(3)$ rotational invariance of the $d=3$ oscillator
is broken down to an $so(2)$ subalgebra.\par
The broad lines of our scheme can be repeated in more general cases, like the deformation of potentials which are no longer quadratic (implying an infinite-dimensional dynamical Lie algebra ${\cal G}_d$) and/or the extension of the abelian twist-deformation to a non-abelian case (e.g., the Jordanian twist \cite{og, tol}). There is a whole list of subtle issues which have to be separately investigated in each such case. We can mention, just as an example, the question whether the symmetry under particle exchange is preserved under a Jordanian twist-deformation. 

\appendix

\section{Detecting deformations through multiparticle operators}

In this Appendix we make some comments about the physical implications, at the operational level, of the twist deformations, focusing on the noncommutative $d=2$ and $d=3$ oscillators discussed in Sections {\bf 3} and {\bf 4} respectively. \par
We were able to recover, within our framework, the energy spectrum of the deformed single-particle Hamiltonian derived in \cite{kijanka, scholtz}. A careful examination of this spectrum shows
a very important feature, namely that it can be reproduced by a linear combination of the undeformed generators (see e.g. formula (\ref{unren2}), where the spectrum for ${\bf H}^{\cal F}$ is reconstructed in terms of ${\bf H}$ and $L$).  This observation brings an important consequence. A measurement of the system which only involves single-particle observables
is not able to detect whether the system is truly deformed or not. The deformation we are dealing
with (the abelian Drinfel'd twist) only makes itself manifest in the multi-particle sector.  The measurement of the two-particle observables is required (and sufficient) to detect whether the system is  deformed or not. Indeed, in the undeformed case the total energy $E_{12}$ of a two-particle system is additive,
given by the sum of the energies $E_1$,$E_2$ of its single-particle components:
\begin{eqnarray}
E_{12}&=&E_1+E_2.
\end{eqnarray}
 On the other hand, in the deformed system, the average $\langle E_{12}\rangle$ of the two-particle total energy, computed in the tensor product of the single-particle eigenvectors,
is given by
\begin{eqnarray}\label{defenergy}
\langle E_{12}\rangle &=& E_1+E_2+\langle {\widehat \Omega_{12}}\rangle.
\end{eqnarray}
The Hermitian operator ${\widehat \Omega_{12}}\neq 0$ in the right hand side (which is not diagonal in the basis given by the tensor product of single-particle energy eigenvectors) can be regarded as an effective interacting term, while $E_1$, $E_2$ are the same as before. We computed here explicitly ${\widehat \Omega_{12}}$
for the non-commutative oscillator (see formula (\ref{omega2}) 
. It is clear at this point that the importance of the (twisted) Hopf algebra framework
cannot be dismissed. It allows to compute the operator ${\widehat \Omega_{12}}$, which in its turn encodes the information about the deformation. \par
It has to be stressed that an arbitrary Hermitian two-particle operator in the right hand side
of (\ref{defenergy}) is not in general admissible. A set of consistency conditions (which are satisfied
by the operator ${\widehat\Omega_{12}}$ that we determined) have to be fulfilled. They include
the symmetry under particle exchange (see Eq. (\ref{deltadef})) and the associativity of the energy
(see Eq. (\ref{intter2})).

\section{A review on Hopf algebras and Drinfel'd twist}

In this appendix we shall briefly review the most important formulas concerning Hopf algebras \cite{{sweedler},{abe},{bal}} and their Drinfel'd twist \cite{{drin85},{drin88}}. 

A \emph{unital associative algebra} over the field $\mathbf{k}$ is a vector space $A$ over $\mathbf{k}$ along with the $\mathbf{k}$-linear maps  $\mu:A\otimes A\rightarrow A$ (multiplication) and  $\eta:\mathbf{k}\rightarrow A$ (unit) satisfying $\mu(\mu(a\otimes b)\otimes c)=\mu(a\otimes\mu(b\otimes c))$, $\forall a,b,c\in A$ (associativity),  and $\eta(1)=\mathbf{1}$ (existence of unit $\mathbf{1}\in A$).

These notions can be dualized: a \emph{coalgebra} over the field $\mathbf{k}$ is a vector space $C$ over $\mathbf{k}$ along with the $\mathbf{k}$-linear maps $\Delta:C\rightarrow C\otimes C$ (coproduct or comultiplication) and $\epsilon:C\rightarrow\mathbf{k}$ (counit) satisfying $\Delta(a_1)\otimes a_2=a_1\otimes\Delta(a_2)$, $\forall a\in C$ (coassociativity), and $\epsilon(\mathbf{1})=1$ (counitarity). Note that the Sweedler notation \cite{sweedler} $\Delta(a)=\sum_i \left(a_1\right)_i\otimes \left(a_2\right)_i \equiv a_1\otimes a_2$  is being employed.

Let now $(H,\mu, \eta)$ be an algebra over $\mathbf{k}$ and $(H, \Delta, \epsilon)$ be a coalgebra over $\mathbf{k}$. We call $(H,\mu,\eta,\Delta,\epsilon)$, or simply $H$, a \emph{bialgebra} if the structures $\mu$ and $\eta$ and the costructures $\Delta$ and $\epsilon$ are compatible, i.e., $\mu$ are $\eta$ coalgebra homomorphisms and  $\Delta$ and $\epsilon$ are algebra homomorphisms.

Take now an algebra $A$ and a coalgebra $C$ and define, for $f,g \in\operatorname{Hom}(C,A)$, the \emph{convolution} $f\ast g=\mu_A(f\otimes g)\Delta_C$. If $H$ is a bialgebra and there exists an element  $S\in \operatorname{Hom}(H,H)$ which is the inverse of the identity with respect to the convolution operation, i.e., $S\ast \mathbf{1}_{\operatorname{Hom}(H,H)}=\mathbf{1}_{\operatorname{Hom}(H,H)}\ast S=\eta_H\circ\epsilon_H$, $(H, \mu, \eta, \Delta, \epsilon, S)$, or simply $H$, is called a \emph{Hopf algebra}. The unique element $S:H\rightarrow H$ is called \emph{antipode} ou \emph{coinverse}. 

The definition of a Hopf algebra can be summarized by the commutativity of the diagram
\[
\centerline{
\xymatrix{
&H\otimes H
\ar[rr]^{S\otimes id}
&& H\otimes H
\ar[rd]^{\mu}
&
\\
H
\ar[ur]^{\Delta}
\ar[rr]^{\epsilon}
\ar[dr]_{\Delta}
&& \mathbf{k}
\ar[rr]^{\eta}
&& H.
\\
&H\otimes H
\ar[rr]^{id\otimes S}
&& H\otimes H
\ar[ru]_{\mu}
&
}
}\]

We will now introduce the notion of Drinfel'd twist \cite{res}. A Hopf algebra $H$ is said to be almost cocommutative if there exists an invertible element $\mathcal{R}\in H\otimes H$ such that $\tau\circ\Delta=\mathcal{R}\Delta\mathcal{R}^{-1}$, where $\tau$ is the flip map. Moreover, an almost cocommutative Hopf algebra is said to be \emph{quasitriangular} if $(\Delta\otimes id)\mathcal{R}=\mathcal{R}_{13}\mathcal{R}_{23}$ and $(id\otimes \Delta)\mathcal{R}=\mathcal{R}_{13}\mathcal{R}_{12}$, where $\mathcal{R}_{12}=\mathcal{R}^\alpha\otimes\mathcal{R}_\alpha\otimes\mathbf{1}$, $\mathcal{R}_{13}=\mathcal{R}^\alpha\otimes\mathbf{1}\otimes\mathcal{R}_\alpha$ and $\mathcal{R}_{23}=\mathbf{1}\otimes\mathcal{R}^\alpha\otimes\mathcal{R}_\alpha$ (we are denoting $ \mathcal{R}=\mathcal{R}^\alpha\otimes\mathcal{R}_\alpha$ and $\mathcal{R}^{-1}=\bar{\cal R}^\alpha\otimes\bar{\cal R}_\alpha$). The element  $\mathcal{R}$ is called \emph{quasitriangular structure} or \emph{universal R-matrix}. Additionally, if $\mathcal{R}_{21}=\mathcal{R}^{-1}$, $H$ is said to be \emph{triangular}. Every cocommutative Hopf algebra is trivially triangular with $\mathcal{R}=\mathbf{1}\otimes \mathbf{1}$. As a consequence of the definition, we have that the quasitriangular structure $\mathcal{R}$ satisfies the \emph{quantum Yang-Baxter equation} $\mathcal{R}_{12}\mathcal{R}_{13}\mathcal{R}_{23}=\mathcal{R}_{23}\mathcal{R}_{13}\mathcal{R}_{12}$, hence its name \emph{universal R-matrix}.

Let now $(H, \mu, \eta, \Delta, \epsilon, S)$ be a cocommutative Hopf algebra and $\mathcal{F}\in H\otimes H$ a counitary 2-cocycle, i.e., $(\mathbf{1}\otimes\mathcal{F})(id\otimes\Delta)\mathcal{F}=(\mathcal{F}\otimes\mathbf{1})(\Delta\otimes id)\mathcal{F}$ and $(\epsilon\otimes id)\mathcal{F}=\mathbf{1}=(id\otimes \epsilon)\mathcal{F}$. We have that $\chi=\mu(id\otimes S)\mathcal{F}$ is an invertible element of $H$ with $\chi^{-1}=\mu(S\otimes id)\mathcal{F}^{-1}$. Now define $\Delta^\mathcal{F}:H\rightarrow H\otimes H$ and  $S^\mathcal{F}:H\rightarrow H$ as $\Delta^{\mathcal{F}} = \mathcal{F}\Delta \mathcal{F}^{-1}$ and $S^\mathcal{F}=\chi S \chi^{-1}$. It can be shown that $(H, \mu, \eta, \Delta^\mathcal{F}, \epsilon, S^\mathcal{F})$ is a triangular Hopf algebra with universal R-matrix given by $\mathcal{R}=\mathcal{F}_{21}\mathcal{F}^{-1}$. We call the twisted Hopf algebra $(H, \mu, \eta, \Delta^\mathcal{F}, \epsilon, S^\mathcal{F})$ simply $H^\mathcal{F}$. The element $\mathcal{F}$ is called a \emph{twist}, and the notation $\mathcal{F}=f^\alpha\otimes f_\alpha$ and $\mathcal{F}^{-1}=\bar{f}^\alpha\otimes \bar{f}_\alpha$ is frequently employed. It should be pointed out that, as an algebra, $H$ is the same as $H^\mathcal{F}$, that is, they are the same vector space and the algebric structures $\mu$ and $\eta$ remain unchanged.

Let us now consider the case of a Lie algebra $\mathfrak{g}$ consisting of generators $\tau_i$ satisfying the commutation relations $[\tau_i,\tau_j]=iC_{ij}^k\tau_k$. It is not a unital associative algebra in general. However, there is a natural construction of a unital associative algebra which contains $\mathfrak{g}$: the universal enveloping algebra $\mathcal{U}(\mathfrak{g})$, which is the quotient $ \mathcal{U}(\mathfrak{g})=T(\mathfrak{g})/I$, where $T(\mathfrak{g})=\bigoplus_{n\geq 0}\mathfrak{g}^{\otimes n}$ is the tensor algebra of $\mathfrak{g}$ and $I$ the ideal generated by all elements of the form $x\otimes y-y\otimes x - [x,y]$. The Poincar\'e-Birkhoff-Witt theorem guarantees that $\mathcal{U}(\mathfrak{g})$ is the algebra of all polynomials of the generators $\tau_i$ modulo the commutation relations. The universal enveloping algebra $\mathcal{U}(\mathfrak{g})$ has a natural Hopf algebra structure if we define the costructures as $\Delta(\tau_i)= \tau_i\otimes \mathbf{1} + \mathbf{1}\otimes \tau_i$ and $\epsilon (\tau_i) = 0$ and the antipode as $S(\tau_i)=-\tau_i$. Therefore, it can undergo Drinfel'd twist following the procedure above, and the resulting twisted Hopf algebra is called $\mathcal{U}^{\cal F}(\mathfrak{g})$. 

It is only natural to ask which is the linear subspace $\mathfrak{g}^{\cal F}\subset \mathcal{U}^{\cal F}(\mathfrak{g})$ analogous to $\mathfrak{g}\subset \mathcal{U}(\mathfrak{g})$. Its elements are called the deformed generators $\tau_i^{\cal F}$. The three conditions for this subspace (see \cite{wor}) are the following: that $\{\tau_i^{\cal F}\}$ form a basis of $\mathfrak{g}^{\cal F}$; the minimal deformation of the Leibniz rule, $\Delta^{\cal F}(\tau_i^{\cal F})=\tau_i^{\cal F}\otimes\mathbf{1}+f_i^j\otimes\tau_j^{\cal F}$, with $f_i^j\in\mathcal{U}(\mathfrak{g})$; and that, under the deformed adjoint action denoted by $[\cdot,\cdot]_{\cal F}$, the structure constants of $\mathfrak{g}$ are reproduced. There is a canonical procedure to obtain $\mathfrak{g}^{\cal F}$ (see \cite{{aschieri},{aschieri2}}). Take as deformed generators $\tau_i^{\cal F}=\bar{f}^\alpha(\tau_i)\bar{f}^\alpha$ with deformed coproduct given by $\Delta^{\cal F}(\tau_i^{\cal F})=\tau_i^{\cal F}\otimes\mathbf{1}+\bar{\cal R}^\alpha\otimes\bar{\cal R}_\alpha(\tau_i^{\cal F})$. The deformed adjoint action is given by $[\tau_i^{\cal F},\tau_j^{\cal F}]_\mathcal{F}=(\tau_i^{\cal F})_1\cdot \tau_j^{\cal F}\cdot S^{\cal F}\left((\tau_i^{\cal F})_2)\right)$. The $\mathfrak{g}^{\cal F}$  thus constructed meets the three requirements above.

{}~
\\{}~
\par {\large{\bf Acknowledgments}}{} ~\\{}~\par
 P.~G.~C. acknowledges financial support from FAPEMIG. B.~C. acknowledges a TWAS-UNESCO associateship appointment 
at CBPF and CNPq for financial support. R.~K. receives a CNPq grant. The work was supported by Edital Universal CNPq Proc. 472903/2008-0.


\end{document}